\documentclass{mem}

\usepackage{natbib}
\usepackage{txfonts}
\usepackage{balance}
\usepackage{graphicx}
\usepackage[a4paper]{hyperref}
\idline{00}{21}

\def\om{\Omega_{\rm bar}}
\def\len{a_{\rm bar}}
\def\lag{R_{\rm CR}}
\def\vpd{{\cal R}}
\def\kin{{\langle V \rangle}}
\def\pin{{\langle X \rangle}}
\begin{document}

\title{Direct measurements of bar pattern speeds}

\subtitle{}

\author{E. M. Corsini}

\offprints{Enrico Maria Corsini; \email{enricomaria.corsini@unipd.it}}
 
\institute{Dipartimento di Astronomia, Universit\`a di Padova, Padova,
  Italy}

\authorrunning{Corsini}

\titlerunning{Direct measurements of bar pattern speeds}

\abstract{
The dynamics of a barred galaxy depends on the angular velocity or
pattern speed of its bar. Indeed, it is related to the location of
corotation where gravitational and centrifugal forces cancel out in
the rest frame of the bar. 
The only direct method for measuring the bar pattern speed is the
Tremaine-Weinberg technique. This method is best suited to the
analysis of the distribution and kinematics of the stellar component
in absence of significant star formation and patchy dust
obscuration. Therefore, it has been mostly used for early-type barred
galaxies. The main sources of uncertainties on the directly-measured
bar pattern speeds are discussed. There are attempts to overcome the
selection bias of the current sample of direct measurements by
extending the application of the Tremaine-Weinberg method to the
gaseous component.
Furthermore, there is a variety of indirect methods which are based on
the analysis of the gas distribution and kinematics. They have been
largely used to measure the bar pattern speed in late-type barred
galaxies.
Nearly all the bars measured with direct and indirect methods end
close to their corotation radius, i.e., they are as rapidly rotating
as they can be.

 }

\maketitle{}

\section{Introduction}

Strong bars are observed in optical images of roughly half of all the
nearby disk galaxies \citep{Marinova2007, Barazza2008, Aguerri2009}.
Therefore bars are a common feature in the central regions of disk
galaxies. Their growth is partly regulated by the exchange of angular
momentum with the stellar disk and the dark matter halo. For this
reason the dynamical evolution of bars can be used to constrain the
content and distribution of dark matter in the inner regions of galaxy
disks \citep[e.g.,][]{Weinberg1985, Bertin1989, Debattista2000}.

The morphology and dynamics of a barred galaxy depend on the angular
velocity or pattern speed of the bar, $\om$. Usually, the bar pattern
speed is parametrized with the bar rotation rate
$\vpd\equiv\lag/\len$. This is the distance-independent ratio between
the corotation radius $\lag$, where the gravitational and centrifugal
forces cancel out in the rest frame of the bar, and the length of bar
semi-major axis $\len$.  The corotation radius is derived from the bar
pattern speed as $\lag = V_{\rm c}/\om$, where $V_{\rm c}$ is the disk
circular velocity.
As far as the value of $\vpd$ is concerned, if $\vpd < 1.0$ the
stellar orbits are elongated perpendicular to the bar and the bar
dissolves.  For this reason, self-consistent bars cannot exist in this
regime. Bars with $\vpd \gtrsim 1.0$ are close to rotating as fast they
can, and there is no a priori reason for $\vpd$ to be significantly
larger than 1.0. 

The value of $\vpd$ can be used to classify bars into fast ($1.0 \leq
\vpd \leq 1.4$) versus slow ($\vpd > 1.4$), with the dividing value at
1.4 by consensus \citep{Debattista2000}.  Note the value of $\vpd$
does not imply a specify value of the pattern speed.

\section{Indirect methods for measuring the bar pattern speed}

A variety of indirect methods has been used to measure the pattern
speed of bars and their corresponding rotation rates.
The identification of rings with the location of Lindblad's resonances
\citep[e.g.,][]{VegaBeltran1998} and analysis of the offset and shape
of dust lanes which trace the location shocks in the gas flows
\citep{Athanassoula1992, Puerari1997} promise to be simple and
physically motivated methods to derive $\om$. But, they are based on
the correct interpretation of morphological features which are often
elusive.
The comparison of the observed gaseous kinematics to dynamical models
of gas flow \citep[e.g.,][]{Lindblad1996} and the comparison of the
observed morphology to the predictions of numerical simulations
\citep[e.g.,][]{Rautiainen2008} combine both kinematic and
photometric information. Furthermore, dynamical models can be applied
to highly-inclined systems. But, both methods usually do allow to put
strong constraints on the error budget and solution uniqueness.

In spite of being model-dependent, all the above methods give
consistent results. The rotation rate of nearly all the measured bars
is consistent to be $1.0 \leq \vpd \leq 1.4$ within the errors
\citep{Elmegreen1996, Rautiainen2008}. However, the sample is biased
toward late-type barred galaxies, because gas-rich systems are required
for this kind of analysis.

\section{The Tremaine-Weinberg method}

\citet[][hereafter TW]{Tremaine1984} suggested a model-independent way
for measuring the bar pattern speed. They showed that $\om$ can be
determined from readily observable quantities for a tracer population
satisfying the continuity equation. It is
\begin{eqnarray}
\lefteqn{\om\, \sin i =} \nonumber \\
& & \frac{\int^{+\infty}_{-\infty}\,h(Y)\,dY
  \int^{+\infty}_{-\infty}\,V(X,Y)\,\Sigma(X,Y)\,dX}
{\int^{+\infty}_{-\infty}\,h(Y)\,dY
  \int^{+\infty}_{-\infty}\,X\,\Sigma(X,Y)\,dX}, 
\label{eq:tw}
\end{eqnarray}
where $(X,Y)$ are the Cartesian coordinates in the sky plane, with the
origin at the disk center and the $X$ and $Y$ axes aligned with the
disk major and minor axes, $h(Y)$ is an arbitrary weight function,
$\Sigma$ and $V$ are the surface brightness and line-of-sight velocity
of the tracer, and $i$ is the disk inclination.
The integration in $X$ ranges over $-\infty \leq X \leq +\infty$, but
integrating over $-X_0 \leq X \leq X_0 $ is sufficient if the disk is
axisymmetric at $|X|\geq X_0$. Although the integration in $Y$ ranges
over $-\infty \leq Y \leq +\infty$, it is actually performed over an
arbitrary range because of $h(Y)$. For example, a weight function
proportional to a delta function $\delta(Y-Y_0)$ corresponds to an
aperture parallel to the disk major axis and offset by a distance
$Y_0$.  This is the case of the slits and pseudo-slits in long-slit
and integral-field spectroscopy, respectively.

\citet{Merrifield1995} refined the TW method. They suitably normalized
both the numerator and denominator of the right-hand side of
Eq. \ref{eq:tw} with the total luminosity in the aperture. Thus, the
TW equation takes the form
\begin{equation}
\om\,\sin i = \frac{\kin}{\pin}, 
\label{eq:mk}
\end{equation}
where 
\begin{equation}
\pin = \frac{\int^{+\infty}_{-\infty}\,h(Y)\,dY
  \int^{+\infty}_{-\infty}\,X\,\Sigma(X,Y)\,dX}
  {\int^{+\infty}_{-\infty}\,h(Y)\,dY
  \int^{+\infty}_{-\infty}\,\Sigma(X,Y)\,dX} 
\label{eq:pin}
\end{equation}
and
\begin{equation}
\kin = \frac{\int^{+\infty}_{-\infty}h(Y)\,dY
  \int^{+\infty}_{-\infty}V(X,Y)\,\Sigma(X,Y)\,dX}
  {\int^{+\infty}_{-\infty}\,h(Y)\,dY
  \int^{+\infty}_{-\infty}\,\Sigma(X,Y)\,dX}
\label{eq:kin}
\end{equation}
are the luminosity-weighted means of the position and line-of-sight
velocity of the tracer, respectively.
Plotting $\kin$ versus $\pin$ for the different apertures produces a
straight line with slope $\om \sin i$, where the inclination of the
galaxy disk is known from the analysis of the surface-brightness
distribution of the galaxy.

The assumption underlying the TW method is that the observed surface
brightness is proportional to the surface density of the tracer, as
for old stellar populations in the absence of significant star
formation and patchy obscuration of dust. Thus, the method has been
successfully applied to absorption-line spectra of early- and
intermediate-type barred galaxies (Table~1).
Extensions of the TW method were proposed to determine the distinct
pattern speeds of two nested bars within a single galaxy
\citep{Corsini2003, Maciejewski2006} and to measure a pattern speed
which may vary arbitrarily with radius \citep[][see also Meidt, this
  volume]{Merrifield2006}.

Usually the presence of shocks, conversion of gas between different
phases, and star formation on short timescales prevent the application
of the TW method to gas. These limitations were explored by
\citet{Rand2004} and \citet{Hernandez2005} with numerical
experiments. Ionized \citep{Hernandez2004, Hernandez2005,
  Emsellem2006, Fathi2007, Fathi2009, Chemin2009, Gabbasov2009},
atomic \citep{Bureau1999}, and molecular gas \citep{Zimmer2004,
  Rand2004} were used to derive the pattern speed of the bar (and/or
spiral arms) in a growing number of intermediate- and late-type barred
galaxies. The agreement between the bar parameters derived from the
gas-based TW method and those obtained from indirect methods and
numerical simulations suggests that although the gas does not obey the
continuity equation, it can be used to derive the bar pattern
speed. Nevertheless, a detailed estimate of the systematic effects due
to departures from continuity and a comprehensive comparison with
stellar-based TW measurements (and corresponding bar rotation rates)
are still missing.
Moreover, TW measurements of the same galaxy based on different
tracers have been not performed yet.
In the rest of this paper, I will focus on the application of the TW
method to absorption-line spectra, while John E. Beckman will review
the results obtained from emission-line spectra in his contribution
(this volume).

\begin{table*}   
\begin{center}   
\caption{Barred galaxies with $\om$ measured by applying TW method 
  to the stellar component}
\begin{small}
\begin{tabular}{llrrrrrl}   
\noalign{\smallskip}  
\hline 
\noalign{\smallskip}  
\multicolumn{1}{c}{Galaxy}  & 
\multicolumn{1}{c}{Morp. Type} &  
\multicolumn{1}{c}{$D$} &  
\multicolumn{1}{c}{$a_{\rm bar}$}  & 
\multicolumn{1}{c}{$\Omega_{\rm bar}$} & 
\multicolumn{1}{c}{$R_{\rm CR}$} &  
\multicolumn{1}{c}{$\cal{R}$} &
\multicolumn{1}{c}{Ref.} \\
\multicolumn{1}{c}{} & 
\multicolumn{1}{c}{} &
\multicolumn{1}{c}{(Mpc)} &    
\multicolumn{1}{c}{(kpc)}  & 
\multicolumn{1}{c}{(km s$^{-1}$ kpc$^{-1}$)} & 
\multicolumn{1}{c}{(kpc)} &
\multicolumn{1}{c}{} &  
\multicolumn{1}{c}{} \\
\multicolumn{1}{c}{(1)} & 
\multicolumn{1}{c}{(2)} &
\multicolumn{1}{c}{(3)} &    
\multicolumn{1}{c}{(4)} & 
\multicolumn{1}{c}{(5)} & 
\multicolumn{1}{c}{(6)} &
\multicolumn{1}{c}{(7)} &  
\multicolumn{1}{c}{(8)} \\
\noalign{\smallskip}  
\hline 
\noalign{\smallskip}  
ESO 139-G09 & (R)SAB0$^0$(rs)   & 71.9 & $5.9^{+2.2}_{-1.0}$ & $61\pm17$ & $5.1^{+1.8}_{-1.1}$  & $0.8^{+0.3}_{-0.2}$ & A$+$03\\ 
ESO 281-G31 & SB0$^0$(rs)       & 70.1 & $3.7\pm0.3$        & $31\pm12$ & $6.8^{+4.1}_{-1.4}$  & $1.8^{+1.1}_{-0.4}$ & G$+$03\\ 
IC 874      & SB0$^0$(rs)       & 34.7 & $3.3^{+0.9}_{-0.8}$ & $42\pm14$ & $4.5^{+2.2}_{-1.1}$  & $1.4^{+0.7}_{-0.4}$ & A$+$03\\ 
NGC 271     & (R$'$)SBab(rs)    & 50.3 & $7.1\pm0.2$        & $32\pm18$ & $10.7^{+7.3}_{-3.9}$ & $1.5^{+1.0}_{-0.5}$ & G$+$03\\ 
NGC 936     & SB0$^+$(rs)       & 14.9 & $3.6\pm0.4$        & $66\pm15$ & $5.0\pm1.1$         & $1.4^{+0.5}_{-0.4}$ & MK95\\
NGC 1023    & SB0$^-$(rs)       &  5.8 & $1.9\pm0.1$        &$181\pm64$ & $1.5^{+1.0}_{-0.6}$  & $0.8^{+0.5}_{-0.3}$ & D$+$02\\ 
NGC 1308    & SB0/a(r)          & 82.4 & $5.0^{+0.7}_{-1.4}$ & $99\pm35$ & $3.6^{+1.8}_{-0.9}$  & $0.8^{+0.4}_{-0.2}$ & A$+$03\\ 
NGC 1358    & SAB0/a(r)         & 51.6 & $4.8\pm0.8$        & $37\pm18$ & $5.8^{+4.8}_{-1.8}$  & $1.2^{+1.0}_{-0.4}$ & G$+$03\\ 
NGC 1440    & (R$'$)SB0$^0$(rs):& 18.4 & $2.2\pm0.5$        & $82\pm19$ & $3.4^{+1.0}_{-0.6}$  & $1.6^{+0.5}_{-0.3}$ & A$+$03\\ 
NGC 2523    & SBbc(r)           & 46.0 & $7.5\pm1.0$        & $30\pm7$  & $9.9^{+3.2}_{-1.9}$  & $1.3^{+0.7}_{-0.5}$ & T$+$07\\ 
NGC 2950    & (R)SB0$^0$(r)     & 19.7 & $3.3\pm0.2$        &$117\pm25$ & $3.1^{+0.8}_{-0.6}$  & $1.0^{+0.3}_{-0.2}$ & C$+$03\\ 
NGC 3412    & SB0$^0$(s)        & 16.0 & $2.4\pm0.2$        & $57\pm16$ & $3.6^{+1.3}_{-0.8}$  & $1.5^{+0.6}_{-0.3}$ & A$+$03\\
NGC 3992    & SBbc(rs)          & 16.4 & $4.5\pm1.0$        & $72\pm5$  & $3.6\pm0.2$         & $0.8\pm0.2$        & G$+$03\\ 
NGC 4245    & SB0/a(r):         & 15.6 & $2.9\pm0.4$        & $62\pm25$ & $3.2^{+2.2}_{-0.9}$  & $1.1^{+1.1}_{-0.4}$ & T$+$07\\ 
NGC 4431    & dSB0/a            & 15.0 & $1.6\pm0.1$        &$102\pm26$ & $0.9^{+0.3}_{-0.2}$  & $0.6^{+0.2}_{-0.1}$ & C$+$07\\ 
NGC 4596    & SB0$^+$(r)        & 29.3 & $7.5\pm1.1$        & $28\pm7$  & $8.6^{+2.8}_{-1.7}$  & $1.1^{+0.7}_{-0.3}$ & G$+$99\\
NGC 7079    & SB0$^0$(s)        & 32.8 & $4.0\pm0.6$        & $53\pm1$  & $4.9\pm0.2$         & $1.2^{+0.3}_{-0.2}$ & DW04\\
\noalign{\smallskip}  
\hline
\noalign{\smallskip}  
\noalign{\smallskip}  
\noalign{\smallskip}  
\end{tabular}
\begin{minipage}{13.4cm}  
NOTE -- 
Col.(2): Morphological classification from \citet[][RC3]{RC3}, except
for ESO 281-G31 (NASA/IPAC Extragalatic Database, NED) and NGC~4431
\citep{Barazza2002}. NGC 2950 is a double-barred galaxy and the listed
values refer to its primary bar.
Col.(3): Distance obtained as $V_{\rm CBR}/H_0$ with $V_{\rm CBR}$
from RC3, except for ESO 281-G31 (NED) and NGC~1308 (NED). The Hubble
constant is assumed to be $H_0=75$ km~s$^{-1}$~Mpc$^{-1}$.
Col.(4): Bar length from reference papers, except for NGC~936
\citep{Kent1989}. Uncertainties of $\pm14\%$ (corresponding to the
median error of the remaining galaxies) are assigned to the bar
lengths of NGC~2523, NGC~4245, and NGC~4596, since errors are not
quoted in the reference papers.
Col.(5): Bar pattern speed from reference papers.
Col.(6): Corotation radius from reference papers. Errors for NGC~2523
and NGC~4245 are not given in \citet{Treuthardt2007}. Uncertainties
are assigned according to the quoted errors on their pattern speeds.
Col.(7): Bar rotation rate from reference papers. Uncertainties of
NGC~936, NGC~2523, NGC~4245, and NGC~4596 are assigned according to
quoted errors on bar length and corotation radius.
Col.(8): Reference papers.
\end{minipage}   
\end{small}  
\end{center}   
\end{table*}

\section{Error budget}

The main sources of uncertainties in TW measurements of $\om$ are
summarized as follows.

\smallskip
\noindent
{\it Centering errors:\/} The value of $\om$ can be significantly
affected by small errors in identifying the position $(X_{\rm
  C},Y_{\rm C})$ of galaxy center and in measuring the value $V_{\rm
  sys}$ of systemic velocity. This effect was already recognized by
\citet{Tremaine1984}. To counter it, they suggested to adopt a weight
function which is odd in $Y$, since barred galaxies are nearly
point-symmetric about their centers.
In long-slit spectroscopy the centering problem translates into one of
fixing an arbitrary reference position and velocity frame common to
all the slits. This in general is a much easier task to achieve. To
this aim \citet{Merrifield1995} rewrote Eq.~\ref{eq:mk} as
\begin{equation}
\om\,\sin i = \frac{\kin-V_{\rm sys}}{\pin-X_{\rm C}}.
\end{equation}
In integral-field spectroscopy the centering errors are minimized by
the unambiguous determination of the common reference frame, which
allows to know the exact position at which velocity and surface
brightness of the tracer are measured.

\smallskip
\noindent
{\it Signal-to-noise ratios:\/} $\kin$ and $\pin$ measure differences
of velocity and luminosity across $X=0$, respectively and are
suscep\-ti\-ble to the noise of available data.
The signal-to-noise ratio of the spectral data can be increased by
collapsing a long-slit spectrum along its spatial direction
\citep{Merrifield1995} or by coadding all the spectra within a
pseudo-slit \citep{Debattista2004}. This produces a single
one-dimensional spectrum with a high signal-to-noise ratio. The mean
Doppler-shift of its absorption lines gives the $\kin$ value, which is
required by the TW method.
Broad-band luminosity profiles have higher signal-to-noise ratios than
luminosity profiles derived from spectra, particularly at large
radii. Therefore, they can be adopted to compute the value of $\pin$
at each position of the slits or pseudo-slits \citep{Aguerri2003}.

\smallskip
\noindent
{\it Uncertainties on the disk position angle:\/} The TW method
requires that the slits (or pseudo-slits) be exactly parallel to the
disk major axis. A careful determination of the disk position angle
PA$_{\rm disk}$ is therefore required before placing the slits or
extracting the pseudo-slits.
The maximum permitted misalignment between the position angle of the
slits (or pseudo-slits) and PA$_{\rm disk}$ to have reliable $\om$
measurements depends on the galaxy inclination and bar orientation
with respect the line of nodes \citep{Debattista2003}. It ranges from
$1^\circ$ to $4^\circ$ for $\Delta \om/\om=0.3$. Galaxies with an
inclination of about $60^\circ$ and a bar at about $20^\circ$ from the
line of nodes on the disk plane are less sensitive to misalignment.
The analysis of the galaxy isophotes mapped by deep imaging of the
axisymmetric region of a disk gives PA$_{\rm disk}$ and $i$. Imaging
is also useful to identify and discard target galaxies with structures
(e.g., outer rings, spiral arms, warped or non-axisymmetric disks)
which may interfere with the accurate measurement of PA$_{\rm disk}$
and $i$ and affect the determination of $\om$.

\smallskip
\noindent
{\it Dust obscuration and star formation:\/} The TW method requires
that the observed surface brightness of the tracer be proportional to
its surface density. This is almost strictly satisfied by old stars in
early-type disk galaxies, because they are characterized by slow star
formation rate and low content of dust.
\citet{Gerssen2007} investigated the effects of dust obscuration and
star formation on the stellar-based TW measurements by means of
numerical simulations.
They find that $\Delta \om/\om = 0.05$ for a diffuse disk of dust with
a typically observed value of extinction $A_V = 3$. A unrealistically
large $A_V=8$ gives $\Delta \om/\om \leq 0.15$.
In addition, barred galaxies often display prominent dust lanes which
run along the leading edges of the bar from the end of the bar toward
the center of the galaxy \citep[see][for a morphological
  classification of dust lanes]{Athanassoula1992}. Dust lanes tend to
increase the TW-derived value of $\om$ when the position angle of the
bar with respect to the disk major axis is PA$_{\rm bar} > 0^\circ$
and decrease it when PA$_{\rm bar} < 0^\circ$. It is $0.08 \leq \Delta
\om/\om \leq 0.25$ for realistic dust lanes with
$A_V\simeq3$. Including star formation does not affect these
conclusions. These experiments show that it is possible to extend the
application of the TW method to the stellar component of late-type
barred galaxies (see also Gerssen \& Debattista, this volume). The
effects of dust obscuration could be further minimized by performing
near-infrared spectroscopy.

\smallskip
\noindent
{\it Number of slits (or pseudo-slits):\/} The value of $\om \sin{i}$
is derived by a straight-line fit to $\pin$ and $\kin$ data. They are
measured along different slits (or pseudo-slits) crossing the galaxy
bar and parallel to the disk major axis. Therefore, the accuracy of
$\om$ determination depends also on the number of the observed slits
(or pseudo-slits).
It ranges from 2 \citep[NGC~1358, $\Delta
  \om/\om=0.49$;][]{Gerssen2003} to 9 (slits for the primary bar of
NGC~2950, $\Delta \om/\om=0.21$; \citealt{Corsini2003}; pseudo-slits
for NGC~7079, $\Delta \om/\om=0.02$; \citealt{Debattista2004}). The
most common case corresponds to 3 slits, they were measured for half
of the sample galaxies. The comparison between $\Delta \om/\om$ of
NGC~2950 and NGC~7079 show that $\Delta \om$ is dominated by the
uncertainties on $\pin$ and $\kin$.
The median relative error on $\om$ is $\Delta \om/\om=0.27$.

\smallskip
TW measurement of $\om$ requires no modeling. However, in the absence
of gas velocities at large radii, the determination of $\vpd$ requires
some modeling to recover the disk circular velocity.
The sources of uncertainties on $\vpd$ are discussed hereafter.

\smallskip
\noindent
{\it Uncertainties on the bar length:\/} Determining the length of a
bar is not a entirely trivial task. This is particularly true for SB0
galaxies, for which there is no spiral structure or star formation
beyond the bar marking its end. Moreover, the presence of a large
bulge complicates further the measurement of $\len$
\citep{Aguerri2005}.
Several methods have been developed to derive the bar length
\citep[see][for a list]{Aguerri2009}.
Methods used to measure $\len$ for the galaxies in Table 1 include:
Fourier decomposition of galaxy light to analyze bar-interbar
intensity ratio \citep{Debattista2002, Aguerri2003, Gerssen2003,
  Corsini2003, Corsini2007, Debattista2004} or phase angle of Fourier
mode $m=2$ \citep{Aguerri2003, Gerssen2003, Corsini2003, Corsini2007},
study of the radial profile of ellipticity \citep{Debattista2002} or
phase angle of the deprojected ellipses which best fit galaxy
isophotes \citep{Debattista2002, Aguerri2003, Corsini2003,
  Corsini2007}, identification of a change in the slope of the
surface-brightness profile along the bar major axis
\citep{Gerssen1999}, visual inspection of galaxy images
\citep{Treuthardt2007}, and decomposition of the surface-brightness
distribution \citep{Kent1989, Kent1990, Aguerri2003, Gerssen2003,
  Corsini2003, Corsini2007}.
The relative error on the bar length for galaxies with at least two
independent measurements is $\Delta \len/\len<0.25$ with a median
$\Delta \len/\len=0.14$. For each galaxy $\Delta \len$ is assumed to
be the average of the error intervals at $68\%$ confidence level.

\smallskip
\noindent
{\it Uncertainties on the corotation radius:\/} The corotation radius
is obtained from the bar pattern speed and disk circular velocity. In
the absence of gas which traces the circular velocity at large radii,
$V_{\rm c}$ is recovered from the observed streaming velocities,
velocity dispersions, and light distribution of the stellar component
by applying the asymmetric drift correction \citep[see][p. 354]{GD2ed}. The
difference between the stellar and circular velocity can be fairly
large in the disks of bright SB0 galaxies ($\Delta V/V=0.2$), where
large stellar velocity dispersions ($\sigma\simeq100$ km~s$^{-1}$) are
observed \citep{Debattista2002, Aguerri2003, Corsini2003}.
Relative error $\Delta V_{\rm c}/ V_{\rm c}$ ranges between 0.05 and
0.2 (with $\Delta V_{\rm c}$ defined as the average of $68\%$ error
intervals). It includes the scatter of the observed velocities in the
flat portion of the stellar rotation curve and variation of the
parameters adopted for the asymmetric drift correction.
Uncertainties on the bar pattern speed and disk circular velocity
translate into a relative error on the corotation radius as large as
$\Delta \lag/\lag=0.57$ \citep[NGC~1358;][]{Gerssen2003}. The median
relative error of the sample is $\Delta \lag/\lag=0.28$ (with $\Delta
\lag$ average of $68\%$ error intervals).

\begin{figure*}[]
%\resizebox{\hsize}{!}{\includegraphics[clip=true]{corsini_f01.ps}}
\begin{center}
\resizebox{10cm}{!}{\includegraphics[clip=true]{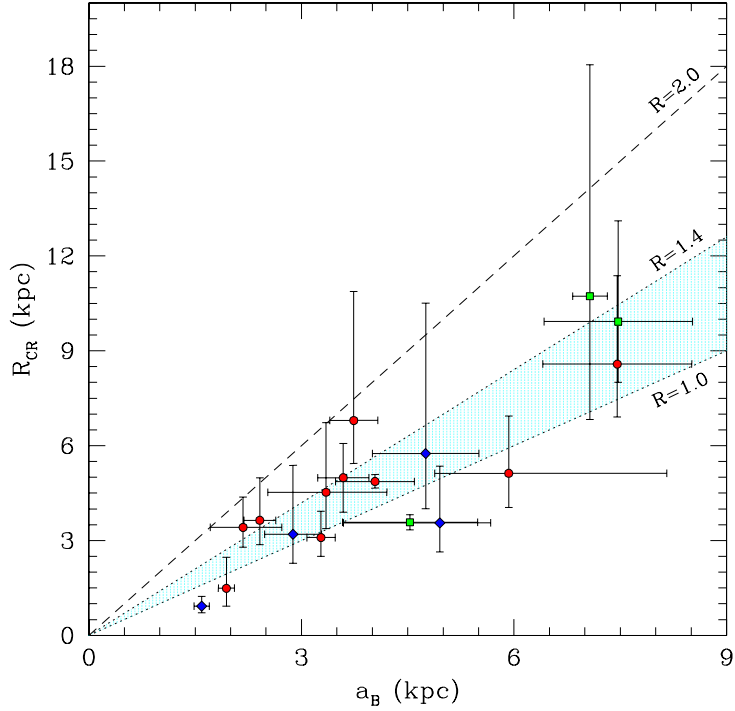}}
\end{center}
\caption{\footnotesize 
  The corotation radius $\lag$ as a function of the bar length $\len$
  for the barred galaxies with $\om$ measured by applying the
  stellar-based TW method. Red circles denote measurements for SB0
  galaxies, blue diamonds correspond to SB0/a galaxies, and green
  squares represent spiral barred galaxies. The dotted lines
  correspond to $\vpd=1$ and $\vpd=1.4$, respectively. They separate
  the forbidden ($\vpd<1$), fast-bar ($1.0\leq\vpd\leq1.4$, hatched
  area), and slow-bar ($\vpd>1.4$) regimes. The dashed line marks
  $\vpd=2$.}
\label{fig:r}
\end{figure*}

\section{Discussion and conclusions}

All the bars in Table~1 are consistent with being fast
(Fig.~\ref{fig:r}). This is particularly true when galaxies with small
uncertainties on $\vpd$ are considered. In fact, the probability that
the bar length is twice as long as the corotation radius ($\vpd>2$) is
$12\%$ for all the sample galaxies and $8\%$ for galaxies with $\Delta
\vpd / \vpd\leq0.3$. The median rotation rate is $\vpd=1.2$.
The fact that some of the values of bar rotation rate are nominally
$\vpd<1$ has been interpreted by \citet{Debattista2003} as due to the
scatter introduced by uncertainties on PA$_{\rm disk}$.
The quoted uncertainties on $\vpd$ are heterogeneous and include both
$68\%$ confidence intervals and maximal errors. It would be very
useful if they were calculated and given in an homogeneous way. For
example, they could be estimated from Monte Carlo simulations based on
the uncertainties on $\len$ and $\lag$.
No trend in $\vpd$ is observed with morphological type. But, the
sample of bars studied so far with the stellar-based TW method is
biased toward the bright and strongly barred SB0 and SB0/a galaxies.
One of them hosts two nested bars (NGC~2950, \citealt{Corsini2003}).
The sample includes only 3 spiral galaxies (NGC~271,
\citealt{Gerssen2003}; NGC~2523, \citealt{Treuthardt2007}; NGC~3992,
\citealt{Gerssen2003}), and one dwarf galaxy (NGC~4431,
\citealt{Corsini2007}).

The bar rotation rate of NGC~4431 is $\vpd = 0.6^{+1.2}_{-0.4}$ at
$99\%$ confidence level \citep{Corsini2007}.  Albeit with large
uncertainty, the probability that the bar ends close to its corotation
radius is about twice as likely as that the bar is much shorter than
the corotation radius. This suggests a common formation mechanism of
the bar in both bright and dwarf galaxies. If their disks were
previously stabilized by massive dark matter halos, bars were not
produced by tidal interactions because they would be slowly rotating
\citep{Noguchi1999}.
But, this is not the case even in the two sample galaxies which show
signs of weak tidal interaction with a close companion (i.e.,
NGC~1023, \citealt{Debattista2002}; NGC~4431, \citealt{Corsini2007}).
There is no difference between TW measurements of the stellar
component in isolated or mildly interacting barred galaxies. Besides,
neither the length nor strength of the bars are found to be correlated
with the local density of the galaxy neighborhoods
\citep{Aguerri2009}.

The bars of ESO~139-G09 \citep{Aguerri2003} and NGC~1358
\citep{Gerssen2003} are weak and fast. Thus, the hypothesis of
\citet{Kormendy1979} that weak bars are the end state of slowed down
fast bars is not supported by observations. They instead favor a
scenario in which weak and strong bars form in the same way.

The $\vpd$ determinations based on the stellar TW method agree with
those obtained by indirect methods, which are largely adopted for
gas-rich galaxies. According to the compilations by
\citet{Elmegreen1996} and \citet{Rautiainen2008}, almost all the
measured bars have $1.0 \leq \vpd \leq 1.4$ within the errors.
The same is true also for the $\vpd$ values obtained from 
pattern speeds measured with the gas-based TW method 
\citep{Bureau1999, Fathi2009, Chemin2009}. A fast bar is ruled 
out by errors only in NGC~2917 \citep[$\vpd>1.7$,][]{Bureau1999} 
and UGC~628 \citep[$\vpd=2.0^{+0.5}_{-0.3}$,][]{Chemin2009}.
If bars of gas-poor lenticulars and early-type spirals have the same
$\vpd$ as gas-rich late-type spirals, then gas is not dynamically
important for the evolution of bar pattern speed
\citep{Debattista2003}. Unfortunately, uncertainties on the measured
$\vpd$ are often not quoted for late-type galaxies. This missing piece
of information is crucial to derive the $\vpd$ distribution as a
function of the morphological type.

Additional work is needed for the stellar TW method to increase both
the accuracy of the $\om$ measurements and extend the number of
studied late-type barred galaxies.
Integral-field spectroscopy overcomes many problems of long-slit
observations and leads to more efficient and accurate TW
measurements. Dwarf barred galaxies are ideal targets to this
aim. They nicely fit the field of view of the available integral-field
spectrographs and are good candidates for testing the predictions that
dark-matter dominated barred galaxies should have a slow bar.
A successful application of the TW method to the stellar component of
late-type barred galaxies would remedy the selection bias present in
the current sample of measured $\om$. This will allow a
straightforward comparison of the results of stellar-based TW method
with indirect and gas-based TW measurements in the same range of
Hubble types.

\begin{acknowledgements}
I would like to thank Victor Debattista for careful reading of the
manuscript and comments that helped to improve it. I would also like
to thank Alfonso Aguerri, John Beckman, and Lorenzo Morelli for their
suggestions.
This research has been made possible by support from Padua University
through grant CPDR095001.
It has made use of the Lyon Extragalactic Database (LEDA) and
NASA/IPAC Extragalactic Database (NED).
\end{acknowledgements}

%%% BIBLIOGRAPHY

\bibliographystyle{aa}

\end{document}